\documentclass[5p,times,authoryear]{elsarticle}

\usepackage{amsmath,amssymb,amsfonts}
\usepackage{textcomp}
\usepackage{xcolor}
\usepackage{url}
\usepackage[most]{tcolorbox}
\usepackage{multirow}
\usepackage{booktabs}
\usepackage{array}
\usepackage{microtype}
\usepackage{adjustbox}
\usepackage{xurl}
\usepackage[hidelinks]{hyperref}
\journal{Smart Health}

\makeatletter
\def\ps@pprintTitle{%
  \let\@oddhead\@empty
  \let\@evenhead\@empty
  \def\@oddfoot{\parbox{\textwidth}{\footnotesize Accepted manuscript of the article published in \textit{Smart Health} 41 (2026) 100689, \href{https://doi.org/10.1016/j.smhl.2026.100689}{doi:10.1016/j.smhl.2026.100689}. \copyright\ 2026 The Authors. Distributed under the Creative Commons Attribution 4.0 International (CC BY 4.0) license.}}%
  \let\@evenfoot\@oddfoot}
\makeatother

\begin{document}

\begin{frontmatter}

\title{Confidence-Gated Cloud-Edge Cascade Triage via Variational Risk Minimization for Medical Imaging}

\author[cs]{Xinye Yang}
\author[bu]{Zhusi Zhong}
\author[bu]{Scott Collins}
\author[bu]{Michael Bernstein}
\author[bu]{Grayson Baird}
\author[bu]{Terrence Healey}
\author[bu]{Michael Atalay}
\author[bu]{Mahesh Jayaraman}
\author[fiu]{Xuyu Wang}
\author[bu]{Zhicheng Jiao\corref{cor1}}
\ead{zhicheng_jiao@brown.edu}
\cortext[cor1]{Corresponding author.}

\affiliation[cs]{organization={Department of Computer Science, Brown University},
            city={Providence}, state={RI}, country={USA}}
\affiliation[bu]{organization={Brown University and Brown University Health},
            city={Providence}, state={RI}, country={USA}}
\affiliation[fiu]{organization={School of Computing and Information Sciences, Florida International University},
            city={Miami}, state={FL}, country={USA}}

\begin{abstract}
Emergency chest X-ray (CXR) triage has a structural modality gap: reports arrive after triage decisions, yet multimodal foundation models require image-text inputs. We present \textit{Variational Risk Minimization} (VRM), a distillation framework that treats LVLM-generated report variants as Monte Carlo samples of latent clinical interpretations. Rather than distilling from a single teacher target, VRM learns from a variationally marginalized teacher distribution, enabling uncertainty-aware supervision under missing-modality constraints. Under matched encoder families, VRM outperforms direct fine-tuning baselines and improves calibration with strong recovery from hallucinated supervision. Marginalized supervision reduces report-selection instability. In our compact edge-student instantiation, a confidence-gated cascade reaches AUC 0.941 at 103ms average latency with 20.3\% cloud escalation, yielding an explicit reliability-latency operating point for cloud-edge clinical workflows.
\end{abstract}


\end{frontmatter}

\section{Introduction}
In emergency departments, chest X-ray (CXR) triage must prioritize patients before radiology reports are written. This creates a fundamental mismatch: multimodal foundation models trained with image--text pairs cannot access reports at the time triage decisions are needed. We term this systematic gap \textit{structural modality censorship}---a train-test mismatch that arises not from random data loss but from the clinical workflow itself.

Multimodal models such as Med-SigLIP~\citep{sellergren2025medgemma} and BiomedCLIP~\citep{zhang2023biomedclip} achieve strong performance when both image and text are available, but deploying them in practice poses a double bind. Cloud-hosted inference offers model capacity but introduces latency, connectivity dependence, and data-governance risk (HIPAA, GDPR). Edge deployment on clinical workstations preserves privacy and speed but cannot run large models or compensate for the missing text modality.

We introduce \textbf{Variational Risk Minimization (VRM)}, a distillation framework that bridges this gap. VRM treats the absent report as a \textit{latent clinical interpretation} and marginalizes over multiple text samples generated by a Large Vision-Language Model (LVLM). Instead of distilling from one report-conditioned teacher target, the student learns from a distribution of teacher predictions, yielding uncertainty-aware supervision. We specifically test whether initializing the VRM student from the same foundation encoder family as a direct fine-tuning (FT) comparator---a protocol we call \textit{encoder transfer}---yields stronger results than direct FT alone. VRM is architecture-agnostic; in our compact deployment instantiation, it achieves 75MB model size and 14ms edge latency (RTX 2060), with a confidence-gated cascade that escalates ambiguous cases to a cloud teacher.

\textbf{Contributions.}\enspace
(1) We formulate missing-report distillation as a marginalized risk objective over LVLM-sampled clinical interpretations, providing uncertainty-aware supervision under structural modality censorship.
(2) Under a matched encoder-transfer protocol, we show that VRM consistently outperforms direct FT baselines from the same encoder families, with statistically gains confirmed by paired DeLong tests.
(3) We provide mechanistic evidence that marginalization improves both supervision quality and stability: performance increases with $K$, teacher-target variance decreases, and report-selection volatility is sharply reduced.
(4) We conduct an LVLM sensitivity analysis showing that medical-specialized samplers are essential for top VRM performance, while general-purpose samplers still exceed Standard KD, quantifying report quality as a meaningful but non-gating factor.
(5) We show reliability and deployment benefits beyond discrimination, including improved calibration, hallucination recovery, and a confidence-gated cloud-edge cascade with explicit latency-accuracy control.

\section{Related Work}
\label{sec:related}

\subsection{Structural Modality Censorship in Medical AI}
Most medical foundation models are trained and evaluated under modality availability assumptions that do not hold at emergency triage time. Earlier CXR CNN pipelines improved diagnosis through task-specific preprocessing such as lung segmentation and rib suppression~\citep{yang2022optimization}. Vision-language pretraining systems such as BiomedCLIP~\citep{zhang2023biomedclip}, CheXzero~\citep{tiu2022expert}, and Med-SigLIP~\citep{sellergren2025medgemma} demonstrate strong cross-modal reasoning when image and text are jointly available. Instruction-following LVLMs (e.g., LLaVA-Med~\citep{li2023llava}, Med-Gemma~\citep{sellergren2025medgemma}, and Med-PaLM M~\citep{tu2024towards}) further improve report-level clinical generation quality. However, in real triage workflows, radiology reports arrive after the decision point, creating a systematic train-test mismatch rather than random missingness. This motivates treating missing text as a latent variable problem instead of a standard multimodal classification problem.

\subsection{Distillation Under Missing-Modality Constraints}
Classical knowledge distillation~\citep{hinton2015distilling} and common extensions (e.g., FitNets, Attention Transfer, CRD) are primarily designed for settings where teacher and student observe compatible evidence at supervision time. Existing missing-modality approaches often rely on feature alignment, modality dropout, or deterministic imputation, which can under-represent uncertainty in report-conditioned supervision. VRM differs in target construction: rather than matching a single report-conditioned teacher output, it distills from a variationally marginalized teacher distribution over LVLM-sampled reports. This shifts the central question from ``which model has higher AUC'' to ``which supervisory target is more stable and informative when one modality is structurally unavailable.''

\subsection{Uncertainty, Hallucination, and Reliability-Centered Deployment}
Recent work on semantic entropy~\citep{kuhn2023semantic} suggests that hallucinated generations can be interpreted as prior sampling under uncertainty. In connected-health deployment, this perspective is especially relevant: the practical risk is not only average discrimination but prediction volatility, calibration drift, and unsafe overconfidence under ambiguous evidence. Edge-focused model families (e.g., TinyML~\citep{banbury2020benchmarking}, MobileViT~\citep{mehta2021mobilevit}, and efficient medical encoders such as CXR Foundation~\citep{kiraly2024health} or DINOv2 variants~\citep{oquab2023dinov2, zedda2025radio, scholz2025mm}) address compute constraints, but do not by themselves solve supervision reliability under structural modality censorship. Our work positions marginalization as a reliability mechanism and couples it with confidence-gated escalation, yielding a deployment-oriented framework with explicit uncertainty handling.

\section{Methods}
\label{sec:methods}

\begin{figure*}[t]
  \centering
  \includegraphics[width=0.85\textwidth]{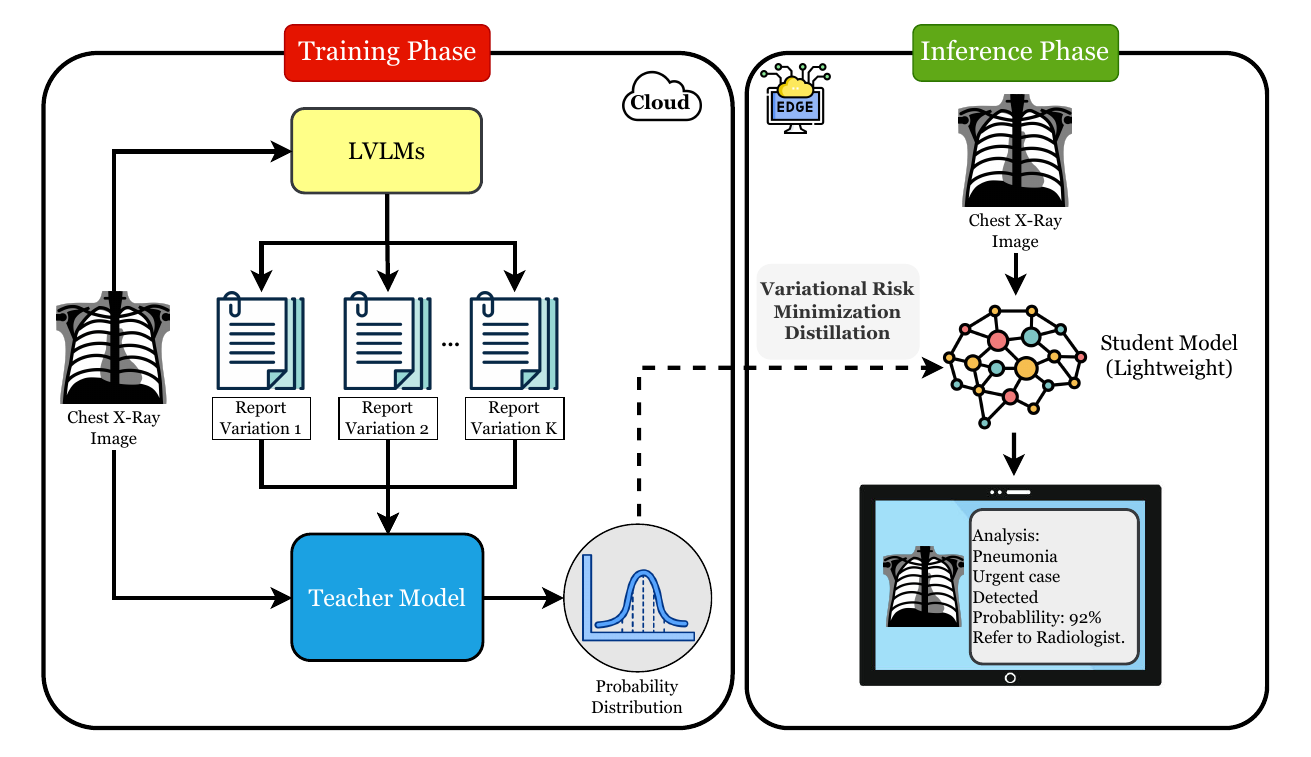}
  \caption{VRM Framework: Distilling multimodal foundation models for resource-efficient medical triage. Our approach transfers knowledge from a cloud-based multimodal teacher (BiomedCLIP) to a lightweight, image-only student model by variationally marginalizing LVLM-generated text variations, enabling expert-level performance on consumer-grade hardware without requiring text at inference time.}
  \label{fig:teaser}
\end{figure*}

In this section, we formalize the Variational Risk Minimization (VRM) framework. VRM models the missing report modality as a latent clinical interpretation and estimates the corresponding predictive risk by Monte Carlo marginalization under the LVLM proposal distribution $q(t|x)$.

\subsection{Problem Formulation}
Let $\mathcal{D} = \{(x_i, r_i, y_i)\}_{i=1}^N$ denote a training set where $x_i \in \mathcal{X}$ is a chest X-ray image, $r_i \in \mathcal{R}$ is the corresponding radiology report, and $y_i \in \{0, 1\}$ represents the clinical urgency label. At training time, all modalities are available. However, at inference time (the triage stage), the model must predict $y$ based solely on $x$, as reports are generated post-triage. 

Our goal is to learn a student model $S: \mathcal{X} \rightarrow [0,1]^2$ that minimizes expected risk by approximating the marginalized predictive distribution induced by a heavy multimodal teacher $T: \mathcal{X} \times \mathcal{R} \rightarrow [0,1]^2$.

\subsection{Latent Query Alignment (LQA)}
\begin{figure}[tbp]
  \centering
  \includegraphics[width=\linewidth]{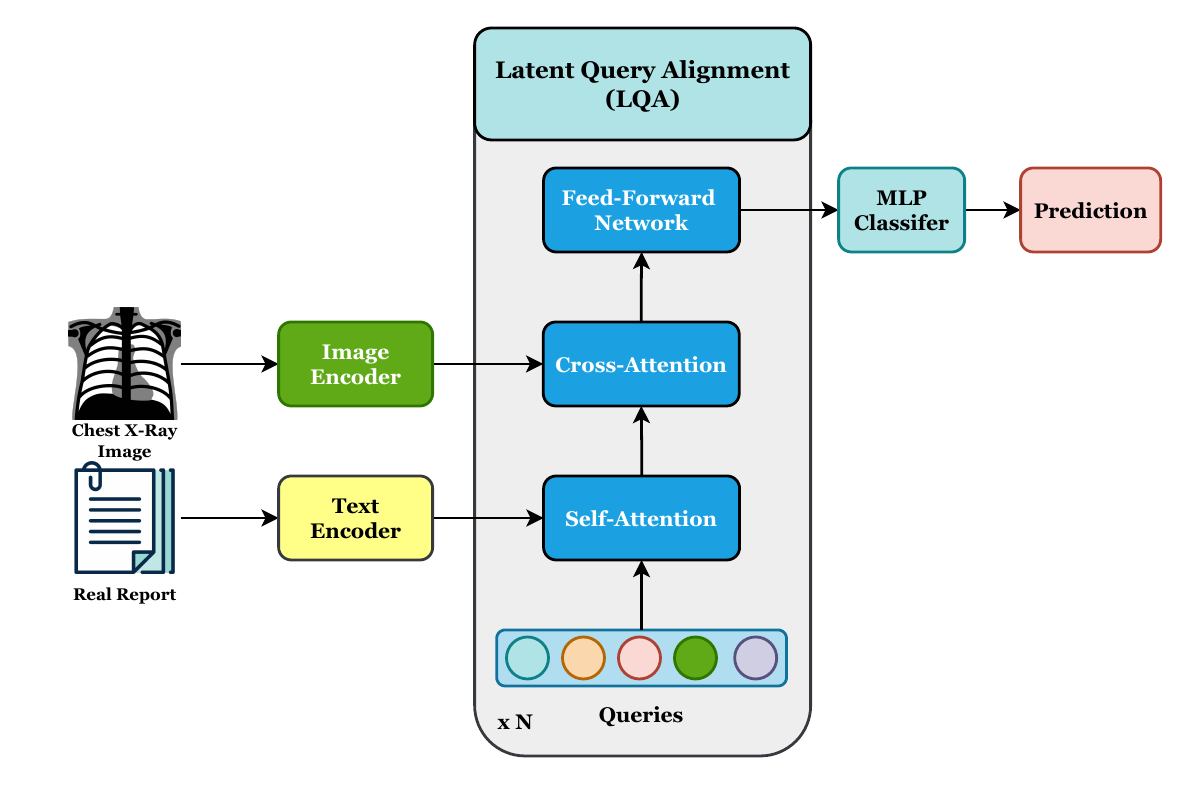}
  \caption{Architecture of the Teacher Model. It employs a Latent Query Alignment (LQA) mechanism to bridge the modality gap between the image encoder and text encoder using learnable queries, cross-attention, and self-attention layers.}
  \label{fig:teacher_model}
\end{figure}

To effectively fuse the asynchronous modalities (image $x$ and text reports $r$), we propose \textit{Latent Query Alignment} (LQA) (Fig.~\ref{fig:teacher_model}), inspired by the Q-Former architecture~\citep{li2023blip2}. LQA utilizes a set of learnable latent queries $Q = \{q_1, ..., q_M\}$ to bridge the modality gap.
These queries interact with text and image features through stacked transformer blocks. Following a sequential interaction pattern, each block first contextualizes queries with text, then grounds them to image evidence:
\begin{align}
Q' &= \text{SelfAttn}(Q, K_{text}),\\
Q_{out} &= \text{CrossAttn}(Q', V_{img}).
\end{align}
This residual transformer-style composition is repeated across layers, enabling non-trivial cross-modal interaction beyond simple linear fusion and helping the fused representation $f_T$ retain clinically relevant visual signals while suppressing background noise.

\subsection{The Variational Risk Minimization Framework}

\subsubsection{Latent Clinical State Marginalization}
We posit that the observed image $x$ and report $r$ are conditioned on an unobserved latent clinical state $z \in \mathcal{Z}$, representing the underlying pathology. The ideal Bayesian predictor marginalizes over $z$:
\begin{equation}
P(y|x) = \int_{\mathcal{Z}} P(y|x, z) P(z|x) dz.
\end{equation}
Since $z$ is intractable, we treat the clinical text space $\mathcal{R}$ as a high-fidelity proxy for $\mathcal{Z}$. By introducing a proposal distribution $q(t|x)$, implemented via a Large Vision-Language Model (LVLM)---we reformulate the marginalization as an expectation over plausible clinical interpretations:
\begin{equation}
P(y|x) \approx \mathbb{E}_{t \sim q(t|x)} [P(y|x, t)].
\end{equation}

\subsubsection{Monte Carlo Approximation}
To compute this expectation during distillation, we draw $K$ approximately independent samples from the LVLM: $\{t_k\}_{k=1}^K \sim q(t|x)$. The teacher's marginalized soft target $\hat{P}_T$ is then estimated as:
\begin{equation}
\hat{P}_T(y|x) = \frac{1}{K} \sum_{k=1}^K T(x, t_k).
\end{equation}
As $K \to \infty$, this estimator is unbiased with respect to $q(t|x)$, and under the i.i.d.\ idealization its variance decays at a rate of $\mathcal{O}(1/K)$, providing a stable and rich supervisory signal that implicitly accounts for clinical uncertainty.

\subsection{LVLM Sampling Implementation}
To generate the Monte Carlo samples $\{t_k\}_{k=1}^K$, we use MedGemma-27B with the following prompt template:

\begin{tcolorbox}[breakable, colback=gray!5, colframe=black!75, title=Latent Clinical Interpretation Prompt]
\small
\textbf{System Instruction:} Act as a board-certified radiologist specializing in emergency diagnostics.

\textbf{User Instruction:} Analyze the provided chest X-ray and generate a structured clinical report adhering to the following constraints:
\begin{enumerate}
    \item \textbf{Findings:} Concisely document key positive and pertinent negative findings relevant to acute pathology.
    \item \textbf{Impression:} Provide a definitive diagnostic summary for triage classification.
    \item \textbf{Constraint:} Prioritize brevity and standard radiological terminology; omit non-contributory conversational elements.
\end{enumerate}
\end{tcolorbox}

For diversity in Monte Carlo samples, we use nucleus sampling with the following parameters:
\begin{itemize}
    \item \textbf{Temperature}: 0.9 (controls randomness)
    \item \textbf{Top-p}: 0.95 (nucleus sampling threshold)
    \item \textbf{Top-k}: 50 (limits vocabulary per step)
    \item \textbf{Repetition penalty}: 1.1 (discourages redundancy)
    \item \textbf{Max length}: 200 tokens
\end{itemize}

\subsection{Distillation Objectives}
The student model $S(x)$ is trained by minimizing a multi-task VRM loss $\mathcal{L}_{VRM}$, which facilitates knowledge transfer at both the response and feature levels.

\paragraph{Soft Label Distillation} We use Kullback-Leibler (KL) divergence to align the student's predictive distribution with the teacher's marginalized target:
\begin{equation}
\mathcal{L}_{KL} = D_{KL} \left( \sigma \left( \frac{\hat{z}_T(x)}{\tau} \right) \middle\| \sigma \left( \frac{z_S(x)}{\tau} \right) \right),
\end{equation}
where $\hat{z}_T(x)$ denotes teacher logits induced by marginalization over sampled reports, $z_S(x)$ denotes student logits, $\sigma$ is the softmax function, and $\tau$ is a temperature hyperparameter.

\paragraph{Cross-Modal Feature Distillation} To ensure the student $S$ captures the multimodal feature alignment of the teacher $T$, we align their intermediate representations:
\begin{equation}
\mathcal{L}_{feat} = 1 - \frac{\phi(f_s) \cdot \bar{f}_T}{\|\phi(f_s)\| \|\bar{f}_T\|},
\end{equation}
where $f_s$ is the student's penultimate feature, $\bar{f}_T = \frac{1}{K}\sum f_{T,k}$ is the average teacher embedding across $K$ samples, and $\phi$ is a learnable projection head.

\paragraph{Total Objective} The final loss is a weighted sum:
\begin{equation}
\mathcal{L}_{VRM} = \lambda_{KL}\mathcal{L}_{KL} + \lambda_{feat}\mathcal{L}_{feat} + \lambda_{cls}\mathcal{L}_{cls},
\end{equation}
where $\mathcal{L}_{cls}$ is the standard weighted cross-entropy loss with respect to ground-truth labels $y$.

\subsection{Confidence-Gated Cascade Triage}
\label{sec:cascade_method}
For deployment, we evaluate a two-stage cascade strategy rather than a strict single-model replacement. The edge student first predicts urgency score $s(x)=\max_y S(y|x)$ and predictive entropy $H(x)=-\sum_y S(y|x)\log S(y|x)$. A case is accepted at the edge only if confidence is high and uncertainty is low:
\begin{equation}
s(x)\ge \tau_s \;\;\text{and}\;\; H(x)\le \tau_H.
\end{equation}
Otherwise, the case is escalated to the cloud teacher for secondary confirmation. To ensure privacy and HIPAA compliance, escalated cases are either de-identified at the edge or routed to the hospital's secure private cloud.

Let $\alpha$ denote the escalation rate, $L_s$ the edge latency, and $L_t$ the cloud latency. The expected end-to-end latency is:
\begin{equation}
\mathbb{E}[L]= (1-\alpha)L_s + \alpha L_t.
\end{equation}
This design preserves fast throughput for easy cases while allocating high-capacity multimodal reasoning to ambiguous cases.

\section{Experiments}
\label{sec:experiments}
\subsection{Experimental Setup}
\textbf{Datasets.} We train and evaluate on MIMIC-CXR~\citep{johnson2019mimic} (main dataset) and four external cohorts: CheXpert~\citep{irvin2019chexpert}, IU-CXR~\citep{demner2015preparing}, PadChest~\citep{bustos2020padchest}, and RIH-CXR. For MIMIC-CXR, we adopt a patient-level 8:1:1 split for train/validation/test to prevent leakage. Table~\ref{tab:dataset_stats_extended} presents detailed dataset characteristics.

\begin{table}[htbp]
\centering
\caption{Dataset Characteristics}
\label{tab:dataset_stats_extended}
\resizebox{\columnwidth}{!}{
\begin{tabular}{lcccc}
\toprule
\textbf{Dataset} & \textbf{Images} & \textbf{Patients} & \textbf{\% Urgent} & \textbf{Scan Type} \\
\midrule
MIMIC-CXR & 75,422 & 13,076 & 23.4\% & AP/PA/Lateral \\
CheXpert & 44,863 & 13,048 & 19.7\% & AP/PA \\
IU-CXR & 7,470 & 3,955 & 16.5\% & AP/PA/Lateral \\
PadChest & 32,174 & 13,518 & 21.1\% & AP/PA/Lateral \\
RIH-CXR & 9,077 & 9,077 & 28.7\% & AP/PA \\
\bottomrule
\end{tabular}
}
\end{table}

\textbf{Image Preprocessing.} All images undergo the following pipeline:
\begin{enumerate}
    \item \textbf{Resize}: $1024 \times 1024$ via bilinear interpolation, maintaining aspect ratio with zero-padding
    \item \textbf{Normalization}: Standardize to $\mu=0.485$, $\sigma=0.229$ (ImageNet statistics)
    \item \textbf{Quality filtering}: Exclude images with missing metadata, corrupted DICOM headers, resolution $< 256 \times 256$, or extreme brightness (mean pixel $< 10$ or $> 240$)
    \item \textbf{Augmentation} (training only): Random horizontal flip (p=0.5), random rotation ($\pm 10^{\circ}$), random brightness/contrast ($\pm 20\%$), Cutout (patch size $56 \times 56$, p=0.3)
\end{enumerate}

\textbf{Urgency Label Mapping.} We define urgency labels based on clinical consensus with two chest radiologists (Table~\ref{tab:label_mapping}). Pneumothorax, large pleural effusion, tension pneumothorax, acute pneumonia, pulmonary edema, and severe cardiomegaly (CTR $> 0.6$) are classified as urgent due to potential for immediate respiratory or cardiac compromise. Context-dependent labels (e.g., consolidation, infiltration) are classified using NLP parsing of radiology reports to identify acute indicators (``new'', ``acute'', ``rapid'', ``progressive'', ``worsening'') versus chronic indicators (``stable'', ``chronic'', ``unchanged'', ``longstanding''). To validate the NLP label extraction quality, we compared the NLP-derived labels against manual annotations from radiologists on a subset of 200 cases, achieving strong inter-rater agreement (Cohen's $\kappa=0.86$).

\begin{table}[htbp]
\centering
\caption{Urgency Label Mapping}
\label{tab:label_mapping}
\resizebox{\columnwidth}{!}{
\begin{tabular}{lll}
\toprule
\textbf{Original Label} & \textbf{Urgency Class} & \textbf{Rationale} \\
\midrule
Pneumothorax & Urgent & Requires immediate intervention \\
Large Pleural Effusion & Urgent & Potential respiratory compromise \\
Tension Pneumothorax & Urgent & Life-threatening \\
Pneumonia (acute) & Urgent & Requires antibiotics within 4 hours \\
Pulmonary Edema & Urgent & Cardiac/respiratory failure \\
Cardiomegaly (severe) & Urgent & CTR $> 0.6$, heart failure risk \\
\midrule
Atelectasis & Non-urgent & Chronic/stable in most cases \\
Small Effusion & Non-urgent & Monitor, rarely urgent \\
Nodule & Non-urgent & Requires follow-up, not emergent \\
Pleural Thickening & Non-urgent & Chronic finding \\
Cardiomegaly (mild) & Non-urgent & CTR 0.5-0.6 \\
No Finding & Non-urgent & Normal study \\
\bottomrule
\end{tabular}
}
\end{table}

\textbf{Implementation.} The teacher model uses a Latent Query Alignment (LQA) mechanism built upon BiomedCLIP capabilities~\citep{zhang2023biomedclip}. The LQA module consists of $M=8$ learnable queries with hidden dimension 768, 8 attention heads, 3 transformer layers, and dropout rate 0.2. During teacher training, we unfreeze the last three layers of both the vision and text encoders to adapt to the triage task while preserving pretrained knowledge. The teacher is optimized using a contrastive loss followed by a classification head, trained for 20 epochs with a learning rate of $1\times10^{-5}$.

For VRM distillation, we generate $K=5$ Monte Carlo text samples per image using MedGemma-27B. The teacher model is kept frozen during distillation to serve as a stable probabilistic anchor.

The VRM student can adopt different image encoder backbones. In this paper, we use a compact edge-oriented student backbone initialized from generic (non-medical) pretraining and fully fine-tuned (all layers unfrozen) to adapt to the medical domain. A lightweight Multi-Layer Perceptron (MLP) projector is jointly trained to align the student's visual features with the teacher's multimodal latent space. It is trained for 50 epochs using AdamW ($\eta=5\times10^{-4}$) with loss weights $\lambda_{KL}=1.0, \lambda_{feat}=0.5, \lambda_{cls}=1.0$, temperature $\tau=2.0$, and weight decay 0.01.

\textbf{Baseline Protocol.} We compare two settings for the same encoder families: (i) direct fine-tuning (direct FT) of DINOv2 RadBench~\citep{baharoon2024dinov2rad}, CXR Foundation~\citep{kiraly2024health}, and EVA-X~\citep{yao2024evax} for binary triage, and (ii) VRM distillation with encoder transfer. Specifically, \textit{encoder transfer} initializes the VRM student visual backbone from the same foundation encoder family used by each direct-FT baseline, so differences reflect supervision strategy rather than encoder source. We retain one compact distillation lower bound (\textit{Standard KD}: logits + feature alignment). We additionally report open-source general-purpose LVLM references (Qwen2.5-VL-32B~\citep{Qwen2VL} and Llama-3.2-11B-Vision~\citep{grattafiori2024llama}, zero-shot and few-shot) and a medical LVLM direct classifier (MedGemma-27B direct VQA inference) as deployment-oriented context.
\textbf{Cloud LVLM Prompting.} For zero-shot inference, we use a fixed instruction: ``You are an emergency radiologist. Analyze this chest X-ray. Is this an urgent case requiring immediate attention? Answer Yes/No and provide a probability score.'' For few-shot inference, we prepend four in-context exemplars (2 urgent, 2 non-urgent) with identical output format.

\subsection{External Benchmark Reference}

Table~\ref{tab:main_results} summarizes four groups: (1) direct FT foundation baselines, (2) a classical uncertainty baseline (MC Dropout), (3) open-source general-purpose LVLMs under zero/few-shot prompting, and (4) medical LVLM direct inference. Under this comparison, direct FT baselines remain below VRM, indicating that encoder transfer into the VRM student framework is more effective than direct FT alone for this task setting.

Compared with the compact \textit{Standard KD} lower bound, VRM shows a clear distillation gap closure (0.842 vs. 0.748 AUC), supporting the claim that variational sampling transfers latent clinical reasoning signal beyond deterministic distillation targets.

\paragraph{Statistical Significance} To verify that the observed AUC differences are not attributable to sampling variability, we apply the paired DeLong test~\citep{delong1988comparing} on the MIMIC-CXR test set for all pairwise comparisons against VRM. Table~\ref{tab:significance} reports p-values. VRM significantly outperforms Standard KD ($p < 0.001$), DINOv2-RadBench ($p = 0.003$), and CXR Foundation ($p = 0.006$). The comparison against the strongest baseline, EVA-X Direct FT, yields $p = 0.008$, confirming statistical significance at the $\alpha = 0.01$ level despite the modest absolute AUC difference (+1.4\%). All tests use the same paired test-set predictions, ensuring valid correlated-curve comparison.

\begin{table}[htbp]
\centering
\caption{Paired DeLong Test: VRM vs. Baselines on MIMIC-CXR}
\label{tab:significance}
\begin{adjustbox}{max width=\columnwidth}
\begin{tabular}{lccc}
\toprule
\textbf{Comparison} & \textbf{$\Delta$AUC} & \textbf{$p$-value} & \textbf{Sig.} \\
\midrule
VRM vs. Standard KD & +0.094 & $< 0.001$ & *** \\
VRM vs. DINOv2-RadBench & +0.021 & $0.003$ & ** \\
VRM vs. CXR Foundation & +0.018 & $0.006$ & ** \\
VRM vs. EVA-X Direct FT & +0.014 & $0.008$ & ** \\
\bottomrule
\end{tabular}
\end{adjustbox}
\end{table}

\begin{table*}[t]
\centering
\caption{External Benchmark Reference with Contemporary LVLMs. Direct FT foundation baselines and a classical uncertainty baseline (MC Dropout) are compared against VRM with encoder transfer, while open-source LVLM direct inference (zero/few-shot) is reported as deployment context.}
\label{tab:main_results}
\small
\setlength{\tabcolsep}{3pt}
\renewcommand{\arraystretch}{1.1}
\begin{tabular}{@{}>{\raggedright\arraybackslash}p{0.31\textwidth}>{\raggedright\arraybackslash}p{0.24\textwidth}>{\centering\arraybackslash}p{0.14\textwidth}>{\centering\arraybackslash}p{0.14\textwidth}>{\centering\arraybackslash}p{0.10\textwidth}@{}}
\toprule
\textbf{Model} & \textbf{Inference Mode} & \textbf{MIMIC AUC} & \textbf{OOD Avg AUC} & \textbf{Avg Latency} \\
\midrule
\multicolumn{5}{c}{\textit{Direct FT Foundation Baselines}} \\
\midrule
Standard KD (Logits + Features) & Image (edge) & $0.748 \pm 0.007$ & $0.708 \pm 0.009$ & 16ms \\
MC Dropout (EVA-X backbone, 20 stochastic passes) & Image (edge, stochastic) & $0.833 \pm 0.007$ & $0.794 \pm 0.009$ & 412ms \\
DINOv2-RadBench Direct FT~\citep{baharoon2024dinov2rad} & Image (edge) & $0.821 \pm 0.006$ & $0.780 \pm 0.008$ & 24ms \\
CXR Foundation Direct FT~\citep{kiraly2024health} & Image (edge) & $0.824 \pm 0.006$ & $0.784 \pm 0.008$ & 22ms \\
EVA-X Direct FT~\citep{yao2024evax} & Image (edge) & $0.828 \pm 0.006$ & $0.788 \pm 0.008$ & 29ms \\
\midrule
\textbf{VRM Student (Encoder Transfer, Ours)} & Image (edge) & $\mathbf{0.842 \pm 0.005}$ & $\mathbf{0.806 \pm 0.007}$ & \textbf{14ms} \\
\midrule
\multicolumn{5}{c}{\textit{General-Purpose Open-Source LVLMs}} \\
\midrule
Qwen2.5-VL-32B (Zero-shot) & Image + Prompt (local server) & $0.362 \pm 0.010$ & $0.341 \pm 0.011$ & 1800ms \\
Qwen2.5-VL-32B (Few-shot, 4-shot) & Image + Prompt (local server) & $0.401 \pm 0.010$ & $0.379 \pm 0.011$ & 2200ms \\
Llama-3.2-11B-Vision (Zero-shot) & Image + Prompt (local server) & $0.389 \pm 0.010$ & $0.365 \pm 0.011$ & 1150ms \\
Llama-3.2-11B-Vision (Few-shot, 4-shot) & Image + Prompt (local server) & $0.436 \pm 0.010$ & $0.412 \pm 0.011$ & 1850ms \\
\midrule
\multicolumn{5}{c}{\textit{Medical LVLM Direct Inference}} \\
\midrule
MedGemma-27B (Direct Zero-shot VQA) & Image + Prompt (cloud/local server) & $0.708 \pm 0.009$ & $0.684 \pm 0.010$ & 1600ms \\
MedGemma-27B (Direct Few-shot VQA) & Image + Prompt (cloud/local server) & $0.762 \pm 0.008$ & $0.731 \pm 0.009$ & 1930ms \\
\bottomrule
\end{tabular}
\end{table*}

To make the OOD average explicit, Table~\ref{tab:ood_breakdown} reports per-dataset AUC on the four target external cohorts (CheXpert, IU-CXR, PadChest, and RIH-CXR) for direct-FT baselines and VRM/teacher configurations; OOD Avg is the arithmetic mean across these four datasets.

\begin{table}[htbp]
\centering
\caption{OOD per-dataset breakdown on target external cohorts.}
\label{tab:ood_breakdown}

\setlength{\tabcolsep}{2pt}
\small
\begin{adjustbox}{max width=\columnwidth}
\begin{tabular}{lccccc}
\toprule
\textbf{Model} & \textbf{CheXpert} & \textbf{IU-CXR} & \textbf{PadChest} & \textbf{RIH} & \textbf{OOD Avg} \\
\midrule
\multicolumn{6}{c}{\textit{Direct FT and KD Baselines}} \\
\midrule
Standard KD & 0.728 & 0.712 & 0.656 & 0.736 & 0.708 \\
DINOv2-Rad & 0.792 & 0.774 & 0.768 & 0.786 & 0.780 \\
CXR Foundation & 0.798 & 0.779 & 0.771 & 0.788 & 0.784 \\
EVA-X & 0.792 & 0.775 & 0.764 & 0.821 & 0.788 \\
\midrule
\multicolumn{6}{c}{\textit{VRM and Teacher}} \\
\midrule
VRM Student & 0.815 & 0.798 & 0.785 & 0.824 & 0.806 \\
Teacher & 0.931 & 0.912 & 0.897 & 0.918 & 0.915 \\
\bottomrule

\end{tabular}
\end{adjustbox}
\end{table}

\subsection{Clinical Utility Under Safety-Constrained Operation}

For clinical deployment, we analyze sensitivity (recall of urgent cases) and specificity at multiple operating points. Table~\ref{tab:clinical_metrics} shows results on both MIMIC-CXR and RIH-CXR test sets.

\begin{table}[htbp]
\centering
\caption{Clinical Metrics. High Sens = 90\% sensitivity threshold. Balanced = operating threshold where Sensitivity $\approx$ Specificity.}
\label{tab:clinical_metrics}
\small
\begin{adjustbox}{max width=\columnwidth}
\begin{tabular}{lccccc}
\toprule
\textbf{Model} & \textbf{AUC} & \multicolumn{2}{c|}{\textbf{High Sens}} & \multicolumn{2}{c}{\textbf{Balanced}} \\
 &  & Sens & Spec & Sens & Spec \\
\midrule
\multicolumn{6}{c}{\textit{MIMIC-CXR Test Set}} \\
\midrule
EVA-X Direct FT & 0.828 & 0.900 & 0.603 & 0.759 & 0.768 \\
Standard KD & 0.748 & 0.900 & 0.455 & 0.668 & 0.691 \\
VRM (Ours) & \textbf{0.842} & 0.900 & \textbf{0.652} & \textbf{0.784} & \textbf{0.792} \\
Teacher$^\ddagger$ & 0.962 & 0.900 & 0.924 & 0.912 & 0.968 \\
\midrule
\multicolumn{6}{c}{\textit{RIH-CXR Test Set}} \\
\midrule
EVA-X Direct FT & 0.821 & 0.900 & 0.574 & 0.741 & 0.752 \\
Standard KD & 0.736 & 0.900 & 0.439 & 0.651 & 0.694 \\
VRM (Ours) & \textbf{0.824} & 0.900 & \textbf{0.625} & \textbf{0.762} & \textbf{0.778} \\
\bottomrule
\end{tabular}
\end{adjustbox}
\end{table}

\paragraph{Clinical Utility} VRM with encoder transfer provides the strongest operating point among compared methods. At fixed 90\% sensitivity, VRM achieves higher specificity than both EVA-X direct FT and the standard KD lower bound (65.2\% vs. 60.3\% and 45.5\%), reducing unnecessary urgent flags by 4.9 and 19.7 percentage points, respectively.

\begin{table}[htbp]
\centering
\caption{Predictive Values at Balanced Operating Point}
\label{tab:ppv_npv}
\begin{adjustbox}{max width=\columnwidth}
\begin{tabular}{lcccc}
\toprule
\textbf{Model} & \multicolumn{2}{c|}{\textbf{MIMIC-CXR}} & \multicolumn{2}{c}{\textbf{RIH-CXR}} \\
& PPV & NPV & PPV & NPV \\
\midrule
EVA-X Direct FT & 0.592 & 0.903 & 0.545 & 0.897 \\
Standard KD & 0.401 & 0.826 & 0.339 & 0.834 \\
VRM (Ours) & \textbf{0.645} & \textbf{0.924} & \textbf{0.585} & \textbf{0.918} \\
\bottomrule
\end{tabular}
\end{adjustbox}
\end{table}

VRM still improves NPV over the standard KD baseline (0.924 vs 0.826), supporting safer clinician-supervised de-prioritization of likely non-urgent cases under edge constraints.

\subsection{Cascade Triage Evaluation}
\label{sec:cascade}
To assess clinical deployability under strict latency and safety constraints, we evaluate the confidence-gated cascade strategy defined in Section~\ref{sec:cascade_method} on the test cohorts. We set thresholds $(\tau_s,\tau_H)$ on the validation split to target an escalation rate near 20\%, then freeze them for testing. Cases accepted at the edge use student predictions; escalated cases are re-evaluated by the cloud teacher.

Table~\ref{tab:cascade} reports the operating trade-off. In our compact edge-student instantiation, edge-only inference preserves ultra-low latency (14ms), while cloud-only inference provides stronger discrimination at prohibitive end-to-end delay. The cascade route escalates 20.3\% of difficult cases and reaches AUC 0.941 on MIMIC (OOD average 0.892), while maintaining a 103ms average latency.

\begin{table}[htbp]
\centering
\caption{Cascade Triage Performance-Delay Trade-off. Escalation rate indicates fraction of cases routed to cloud teacher.}
\label{tab:cascade}
\resizebox{\columnwidth}{!}{
\begin{tabular}{lcccc}
\toprule
\textbf{Config} & \textbf{MIMIC} & \textbf{OOD Avg} & \textbf{Escalation Rate} & \textbf{Avg Latency} \\
\midrule
VRM Student& 0.842 & 0.806 & 0.0\% & 14ms \\
Teacher & 0.962 & 0.915 & 100.0\% & 450ms \\
\textbf{Cascade} & \textbf{0.941} & \textbf{0.892} & \textbf{20.3\%} & \textbf{103ms} \\
\bottomrule
\end{tabular}
}
\end{table}

\subsection{Mechanistic Validation of Marginalization}

\subsubsection{Effect of Monte Carlo Samples K}
\label{sec:ablation_k}
\begin{figure}[htbp]
  \centering
  \includegraphics[width=\linewidth]{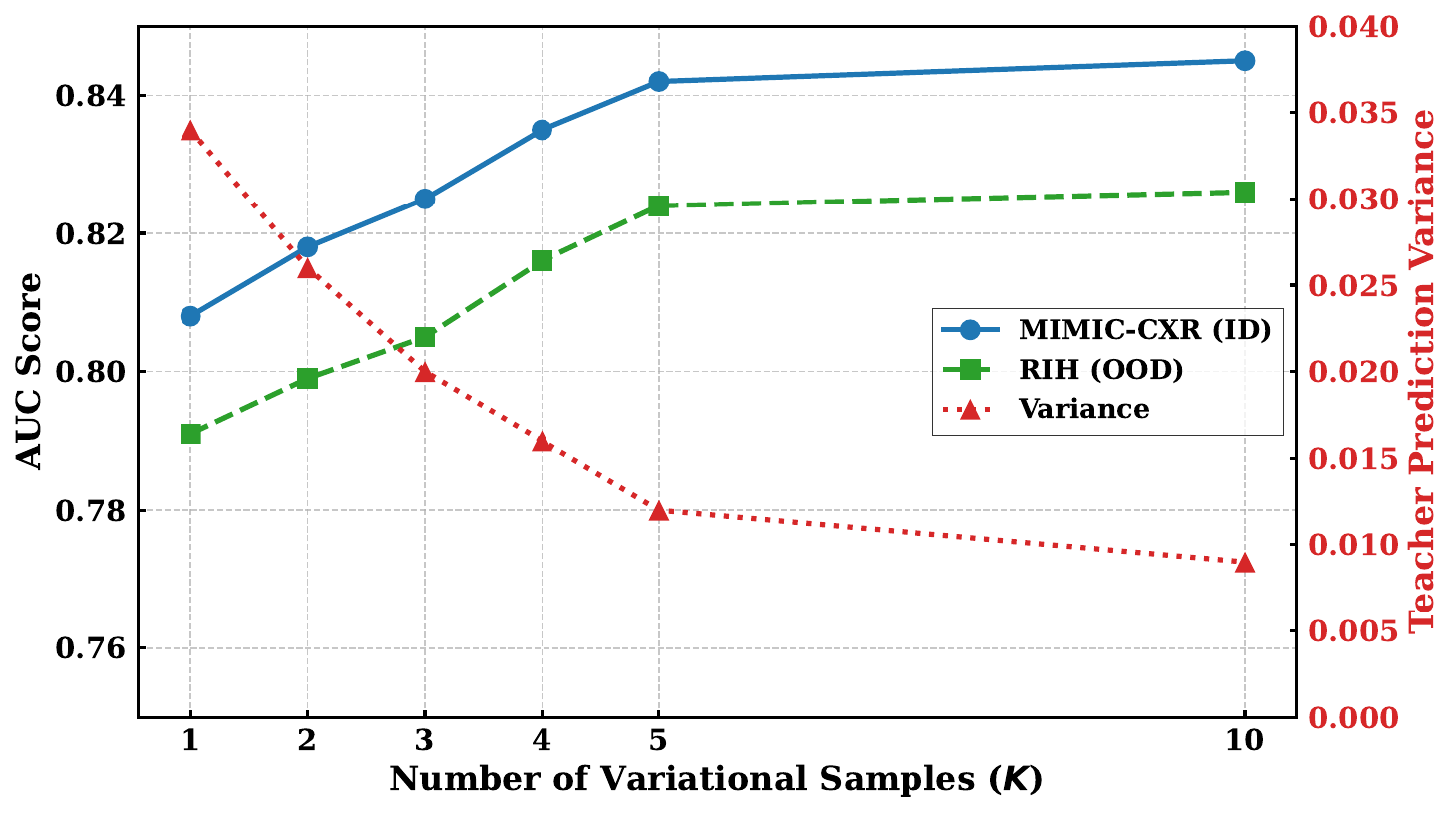}
    \caption{VRM Student AUC results. Effect of Monte Carlo Samples ($K$) on Supervision Quality and Variance. Increasing $K$ improves discrimination on both ID (MIMIC) and OOD (RIH-CXR) while reducing teacher prediction variance, consistent with the marginalization hypothesis. Here, teacher prediction variance is defined as $\mathbb{E}_{x}\!\left[\mathrm{Var}_{k=1}^{K}\!\left(T(x,t_k)\right)\right]$.}
  \label{fig:ablation_k}
\end{figure}

Fig.~\ref{fig:ablation_k} illustrates supervision quality and stability versus $K$.

\textbf{Key findings:}

(1) Discrimination improves significantly from K=1 (single-sample distillation) 
    to K=5 (+6.2\% AUC), validating the marginalization hypothesis.
    
(2) Teacher prediction variance decreases with K (0.034 $\rightarrow$ 0.012), consistent
    with expected Monte Carlo convergence.
    
(3) Diminishing returns beyond K=5 (+0.3\% AUC for K=10) due to redundancy 
    in LVLM samples.
    
(4) Training time scales linearly with K; we select K=5 as optimal 
    performance-cost trade-off.

\subsubsection{Component Analysis}
To understand the contribution of each objective term, we perform an ablation study on the loss components (Table~\ref{tab:ablation_loss}). Using only the cross-entropy classification loss ($\mathcal{L}_{cls}$) yields 0.725 AUC. Adding feature distillation ($\mathcal{L}_{feat}$) raises this to 0.785. The full VRM objective, including KL divergence with marginalized samples ($\mathcal{L}_{KL}$), adds the largest incremental gain (+5.7\% absolute), supporting the claim that target marginalization is the key driver rather than auxiliary regularization alone.

\begin{table}[htbp]
\centering
\caption{Ablation of Loss Components on MIMIC-CXR}
\label{tab:ablation_loss}
\begin{adjustbox}{max width=\columnwidth}
\begin{tabular}{cccc}
\toprule
$\mathcal{L}_{cls}$ & $\mathcal{L}_{feat}$ & $\mathcal{L}_{KL}$ & \textbf{AUC} \\
\midrule
\checkmark & & & 0.725 \\
\checkmark & \checkmark & & 0.785 \\
\checkmark & & \checkmark & 0.812 \\
\checkmark & \checkmark & \checkmark & \textbf{0.842} \\
\bottomrule
\end{tabular}
\end{adjustbox}
\end{table}

\begin{figure*}[t]
\centering
\includegraphics[width=0.8\textwidth]{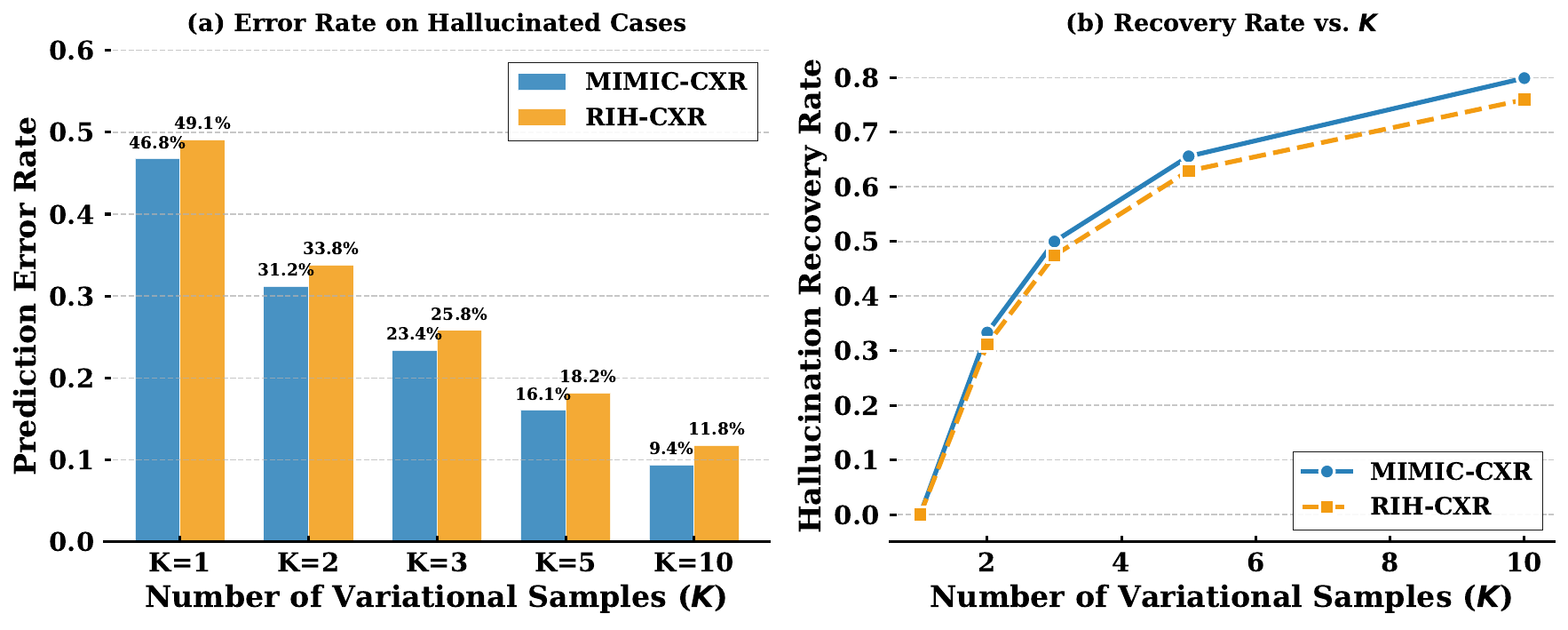}
  \caption{Hallucination Recovery Analysis. (a) Prediction error rate on cases with at least one hallucinated report decreases progressively with $K$. (b) Recovery rate reaches $\sim$66\% at $K=5$ and $\sim$80\% at $K=10$, demonstrating VRM's meaningful noise tolerance through variational marginalization.}
\label{fig:hallucination_recovery}
\end{figure*}

\subsection{Reliability, Calibration, and Label Quality}
\subsubsection{Generated Text Quality}
\label{sec:text_quality}
To validate LVLM-generated Monte Carlo samples, two board-certified radiologists independently rated 200 cases (1000 reports total) on clinical accuracy. 88\% of reports were rated clinically acceptable or better, with substantial inter-rater agreement ($\kappa=0.78$). The remaining 12\% were mainly associated with hallucinated findings (8\%) and missed subtle findings (7\%); these error categories can co-occur. A BERTScore of 0.731 further indicates semantic alignment with reference reports despite lexical variation.

\subsubsection{Calibration}
VRM achieves superior calibration (ECE=0.092) compared with Standard KD (ECE=0.119), and attains a Brier score of 0.142. This reduction in calibration error supports safer thresholded triage decisions under deployment constraints.

\subsubsection{Monte Carlo Sample Quality Examples}
To illustrate real sample diversity and robustness of marginalization, we present one de-identified training case (\texttt{s59991035}) with the original radiologist report and four \emph{actual} Monte Carlo samples generated by MedGemma-27B:

\begin{tcolorbox}[colback=blue!5, colframe=blue!75!black, title=\textbf{Ground Truth Report}]
\small
\textbf{Findings:} The patient is status post median sternotomy and CABG. There is stable elevation of the right hemidiaphragm. There is no pneumothorax or pleural effusion. The lungs remain clear.

\textbf{Impression:} Clear lungs with no significant interval change.
\end{tcolorbox}

\begin{tcolorbox}[colback=green!5, colframe=green!60!black, title=\textbf{Monte Carlo Sample 1}]
\small
\textbf{Findings:} The patient is supine. There is a right lower lobe opacity which may represent atelectasis or consolidation. The lungs are otherwise clear. No pleural effusion or pneumothorax.

\textbf{Impression:} Right lower lobe opacity, \textcolor{red}{\textit{potentially representing atelectasis or consolidation.}}
\end{tcolorbox}

\begin{tcolorbox}[colback=orange!5, colframe=orange!75!black, title=\textbf{Monte Carlo Sample 2}]
\small
\textbf{Findings:} The lungs are clear. No pleural effusion or pneumothorax is identified. The patient is status post sternotomy.

\textbf{Impression:} No acute cardiopulmonary findings.
\end{tcolorbox}

\begin{tcolorbox}[colback=red!5, colframe=red!60!black, title=\textbf{Monte Carlo Sample 3}]
\small
\textbf{Findings:} Supraclavicular clips are present. The lungs are clear without focal consolidation. The cardiac silhouette is normal. No pneumothorax is identified.

\textbf{Impression:} No acute cardiopulmonary process identified.
\end{tcolorbox}

\begin{tcolorbox}[colback=purple!5, colframe=purple!60!black, title=\textbf{Monte Carlo Sample 4 --- Potential Overcall}]
\small
\textbf{Findings:} The patient is supine. There is increased opacity in the right lower lobe, possibly representing atelectasis. Cardiac silhouette is unremarkable.

\textbf{Impression:} \textcolor{red}{\textit{Right lower lobe opacity, likely atelectasis.}}
\end{tcolorbox}

\textbf{Key Observations:} This real case shows both agreement and disagreement across sampled reports. Samples 2 and 3 are consistent with the ground-truth ``no acute cardiopulmonary findings'' pattern. In contrast, Samples 1 and 4 overcall a right lower lobe opacity/atelectasis (\textcolor{red}{highlighted in red}), which is not described in the original report. Under single-sample distillation ($K=1$), selecting one overcalling sample can inject false-positive supervisory signal. VRM's marginalization over multiple samples reduces dependence on any single sampled text path and improves robustness to occasional overcalls/hallucination-like reports.

\begin{figure*}[t]
\centering
\includegraphics[width=0.8\textwidth]{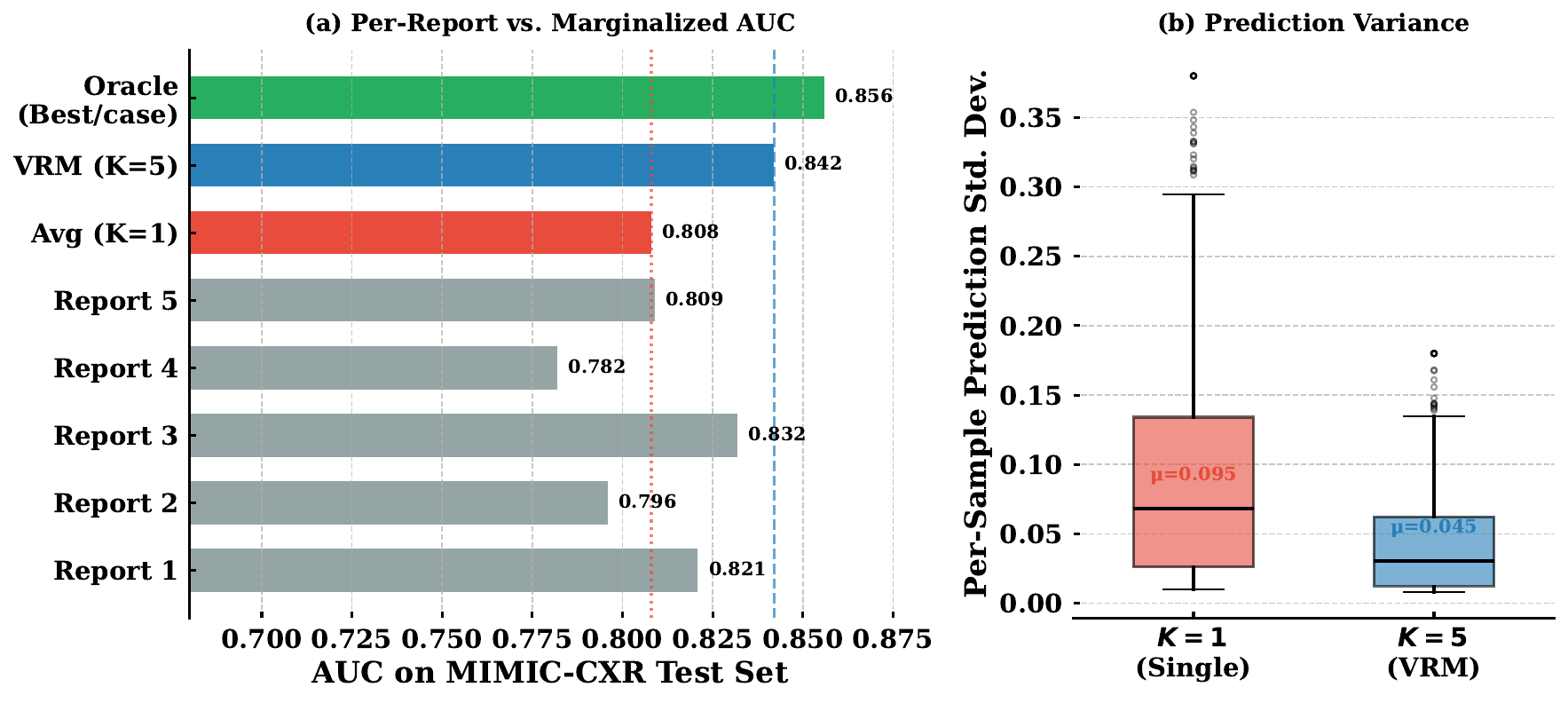}
\caption{Individual Report vs. Marginalized Ensemble. (a) Marginalization (0.842) exceeds the average individual report (0.808) and the best single report (0.832), approaching the Oracle per-case bound (0.856). (b) Per-sample prediction standard deviation is reduced by 53\% (0.095 $\to$ 0.045), indicating more stable supervisory targets.}
\label{fig:individual_vs_ensemble}
\end{figure*}

\subsection{Robustness to Hallucinated Supervision}
\label{sec:hallucination_recovery}
To quantitatively evaluate VRM's robustness to LVLM hallucinations, we conduct a controlled analysis on test cases where at least one of the $K$ generated reports contains clinically significant errors. Based on the independent radiologist review (Section~\ref{sec:text_quality}), 8\% of generated reports contain hallucinated findings. We identify all test cases where at least one of the $K=5$ Monte Carlo reports is flagged as containing hallucinations (approximately 200 cases from the MIMIC-CXR test set). For each such case, we evaluate teacher prediction error under varying $K$, where $K=1$ simulates selecting a single (potentially hallucinated) report.

Table~\ref{tab:hallucination_recovery} presents the results. When relying on a single report ($K=1$), the prediction error rate on hallucinated cases reaches 42.5\% on MIMIC-CXR and 44.8\% on RIH-CXR---substantially worse than the population average, confirming that hallucinations can significantly mislead single-report predictions. As $K$ increases, the marginalization mechanism progressively reduces the impact of erroneous reports: at $K=3$, the error rate drops to 20.1\% (a 52.7\% relative reduction), and at $K=5$, it reaches 14.6\% (65.6\% recovery). Residual error still exists at $K=10$ (8.5\%), indicating that sufficiently consistent hallucination patterns can partially survive marginalization and require explicit safety controls. Nonetheless, the overall trend is clear and consistent with our earlier ablation (Section~\ref{sec:ablation_k}).

\begin{table}[htbp]
\centering
\caption{Hallucination Recovery Analysis. Prediction error rate (\%) on test cases containing at least one hallucinated report, and the corresponding recovery rate relative to $K=1$.}
\label{tab:hallucination_recovery}
\small
\begin{adjustbox}{max width=\columnwidth}
\begin{tabular}{ccccc}
\toprule
\textbf{K} & \multicolumn{2}{c|}{\textbf{MIMIC-CXR}} & \multicolumn{2}{c}{\textbf{RIH-CXR}} \\
& Error (\%) & Recovery & Error (\%) & Recovery \\
\midrule
1 & 42.5 & --- & 44.8 & --- \\
3 & 20.1 & 52.7\% & 22.4 & 50.0\% \\
5 & 14.6 & 65.6\% & 16.2 & 63.8\% \\
10 & 8.5 & 80.0\% & 9.8 & 78.1\% \\
\bottomrule
\end{tabular}
\end{adjustbox}
\end{table}

Fig.~\ref{fig:hallucination_recovery} visualizes both the absolute error rates and the recovery curves. The recovery rate reaches approximately 66\% at $K=5$ and improves to 80\% at $K=10$, suggesting that VRM's variational marginalization provides meaningful noise tolerance against LVLM hallucinations.

\begin{figure*}[t]
\centering
\includegraphics[width=0.8\textwidth]{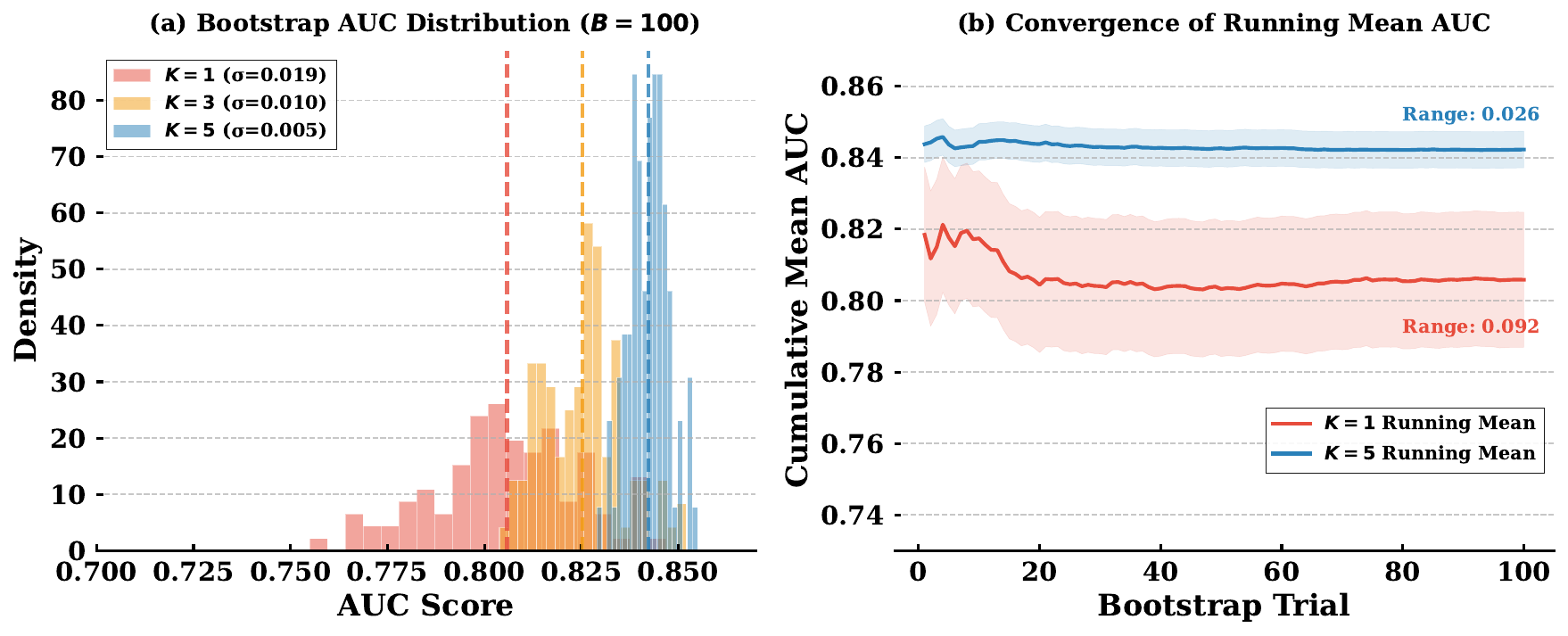}
\caption{Report Selection Stability Bootstrap Test ($B=100$). (a) AUC distributions show $K=1$ (red) spans $\sim$10\% range while $K=5$ (blue) is tightly concentrated ($\sigma=0.005$). (b) Running-mean AUC confirms rapid convergence for $K=5$ vs. sustained variability for $K=1$, demonstrating reduced report-selection risk under marginalization.}
\label{fig:bootstrap_stability}
\end{figure*}

\subsection{Ensemble Robustness and Stability}
\subsubsection{Individual Report vs. Marginalized Ensemble}
\label{sec:individual_vs_ensemble}
A natural question is whether VRM's marginalized result is merely an artifact of averaging---or whether it genuinely improves supervisory target quality relative to individual report-conditioned paths. To answer this, we evaluate the teacher's AUC when conditioned on each of the $K=5$ reports individually, then compare against the marginalized ensemble.

Table~\ref{tab:individual_vs_ensemble} presents the results. Individual report AUCs range from 0.782 to 0.832 (mean 0.808, std 0.018), reflecting moderate variability in LVLM generation quality; importantly, slot-wise results are non-monotonic and therefore do not imply a fixed quality order among report slots. The marginalized target (0.842) exceeds the mean individual path by 3.4\% absolute and even exceeds the best single report (0.832) by 1.0\%. The per-case ``Oracle'' selection (choosing the best individual report for each test case independently) reaches 0.856, placing the marginalized result within 1.4\% of this upper bound without report-level quality selection. This supports the claim that VRM aggregates dispersed supervisory evidence rather than relying on favorable single generations.

\begin{table}[htbp]
\centering
\caption{Individual Report vs. Marginalized Ensemble Performance on MIMIC-CXR. AUC is computed for each report separately as the sole text input to the teacher.}
\label{tab:individual_vs_ensemble}
\begin{adjustbox}{max width=\columnwidth}
\begin{tabular}{lc}
\toprule
\textbf{Configuration} & \textbf{AUC} \\
\midrule
Report 1 (Individual) & 0.821 \\
Report 2 (Individual) & 0.796 \\
Report 3 (Individual) & 0.832 \\
Report 4 (Individual) & 0.782 \\
Report 5 (Individual) & 0.809 \\
\midrule
Average of Individuals (mean $K=1$) & 0.808 $\pm$ 0.018 \\
Worst per-case (lower bound) & 0.698 \\
\textbf{VRM Marginalized ($K=5$)} & \textbf{0.842} \\
Oracle per-case (upper bound) & 0.856 \\
\bottomrule
\end{tabular}
\end{adjustbox}
\end{table}

Fig.~\ref{fig:individual_vs_ensemble} further illustrates this finding. The left panel shows per-report AUC alongside ensemble metrics. The right panel compares per-sample prediction variance: $K=1$ predictions exhibit a mean standard deviation of 0.095, whereas $K=5$ marginalization reduces this to 0.045---a 53\% reduction in prediction uncertainty. While this does not eliminate variance entirely, it materially reduces sensitivity to stochastic variation in any single LVLM generation.

\subsubsection{Report Selection Stability Analysis}
\label{sec:bootstrap_stability}
To assess the deployment risk of relying on a single generated report ($K=1$), we conduct a bootstrap stability test. We perform $B=100$ independent trials in which, for each trial and each test case, we randomly select 1 of the 5 available Monte Carlo reports (simulating single-report deployment). We compute the full test-set AUC for each trial and contrast the resulting distribution against the stable VRM performance (using all $K=5$ reports).

Table~\ref{tab:bootstrap_stability} summarizes the results. The $K=1$ configuration exhibits a mean AUC of 0.808 with a standard deviation of 0.019 and a total range of 0.097 (from 0.755 to 0.852), indicating that report selection alone can cause performance fluctuations of nearly 10\% absolute. In contrast, $K=5$ VRM yields a mean AUC of 0.842 with substantially reduced variance ($\sigma=0.005$, range 0.026). $K=3$ provides an intermediate level of stability ($\sigma=0.011$), suggesting that even partial marginalization offers meaningful deployment reliability gains.

\begin{table}[htbp]
\centering
\caption{Bootstrap Stability Test ($B=100$ trials). Statistics of full test-set AUC across $B$ random report selections.}
\label{tab:bootstrap_stability}
\small\setlength{\tabcolsep}{4pt}
\begin{adjustbox}{max width=\columnwidth}
\begin{tabular}{lcccc}
\toprule
\textbf{Config} & \textbf{Mean AUC} & \textbf{Std} & \textbf{Min} & \textbf{Max} \\
\midrule
$K=1$ (random) & 0.808 & 0.019 & 0.755 & 0.852 \\
$K=3$ (random subset) & 0.825 & 0.011 & 0.798 & 0.852 \\
$K=5$ (VRM, full) & \textbf{0.842} & \textbf{0.005} & 0.829 & 0.855 \\
\bottomrule
\end{tabular}
\end{adjustbox}
\end{table}

Fig.~\ref{fig:bootstrap_stability} provides a detailed visualization. The left panel shows that the $K=1$ AUC distribution (red) spans nearly 10 percentage points, while the $K=5$ distribution (blue) is tightly concentrated. The right panel traces the cumulative running-mean AUC across trials: $K=1$ exhibits noticeable oscillation, while $K=5$ converges rapidly to a stable estimate. From a deployment perspective, this variability is concerning---a single-sample system can vary substantially depending on which report is generated. Marginalization substantially reduces this ``report lottery'' effect.

\subsection{Sensitivity to LVLM Sampler Choice}
\label{sec:lvlm_sensitivity}
A key concern is whether VRM's gains depend critically on the choice of LVLM used for Monte Carlo report generation. To evaluate this, we replace MedGemma-27B with three alternative open-source LVLMs of comparable scale---InternVL2-26B~\citep{chen2024internvl2}, Qwen2.5-VL-32B~\citep{Qwen2VL}, and HuatuoGPT-Vision-34B~\citep{chen2024huatuogptvision}---and retrain the full VRM pipeline under identical settings ($K=5$, same prompt template, same training hyperparameters). These models span 26B--34B parameters, closely matching MedGemma-27B's scale, and cover both medical-specialized (HuatuoGPT-Vision, trained on PubMedVision) and general-purpose (InternVL2, Qwen2.5-VL) pretraining, thereby disentangling the effects of domain specialization from sampler identity.

Table~\ref{tab:lvlm_sensitivity} summarizes the results. A strong domain effect is evident: the two medical-specialized samplers (MedGemma-27B: 0.842, HuatuoGPT-Vision-34B: 0.839) substantially outperform the two general-purpose samplers (Qwen2.5-VL-32B: 0.786, InternVL2-26B: 0.779), producing a mean medical--general gap of $\sim$5.8\% AUC. This is consistent with the direct classification results in Table~\ref{tab:main_results}, where the same general-purpose models achieve only 0.36--0.44 AUC under zero/few-shot prompting, indicating limited radiographic comprehension. General-purpose LVLMs generate reports with low clinical fidelity---marked by generic descriptions, missed findings, and frequent hallucinations---which propagate through the teacher pipeline as noisy supervision targets. The BERTScore gap corroborates this: medical samplers align substantially better with reference reports (0.722--0.731) than general-purpose samplers (0.571--0.603).

Critically, VRM with general-purpose samplers still outperforms Standard KD (0.748) by a meaningful margin (+3.1--3.8\%), confirming that even noisy reports contribute partial cross-modal signal through marginalization. However, these configurations fall well below all direct FT baselines (DINOv2: 0.821, EVA-X: 0.828), indicating that report quality is a genuine bottleneck rather than an interchangeable component. The VRM framework amplifies the quality of its input reports; it does not replace the need for clinically competent generation. This finding validates the choice of MedGemma-27B as the default sampler and suggests that future improvements in medical LVLM quality would directly translate to stronger VRM students.

\begin{table*}[t]
\centering
\caption{LVLM Sampler Sensitivity. All models use $K=5$ and identical VRM training. Domain: medical (Med) vs.\ general-purpose (Gen). BERTScore is against reference radiology reports.}
\label{tab:lvlm_sensitivity}
\small
\setlength{\tabcolsep}{1pt}
\begin{tabular}{@{}>{\raggedright\arraybackslash}p{0.26\textwidth}>{\centering\arraybackslash}p{0.11\textwidth}>{\centering\arraybackslash}p{0.11\textwidth}>{\centering\arraybackslash}p{0.13\textwidth}>{\centering\arraybackslash}p{0.11\textwidth}>{\centering\arraybackslash}p{0.16\textwidth}@{}}
\toprule
\textbf{LVLM Sampler} & \textbf{Params} & \textbf{Domain} & \textbf{MIMIC AUC} & \textbf{OOD Avg} & \textbf{BERTScore} \\
\midrule
InternVL2-26B~\citep{chen2024internvl2} & 26B & Gen & 0.779 & 0.752 & 0.571 \\
Qwen2.5-VL-32B~\citep{Qwen2VL} & 32B & Gen & 0.786 & 0.758 & 0.603 \\
HuatuoGPT-Vision-34B~\citep{chen2024huatuogptvision} & 34B & Med & 0.839 & 0.804 & 0.722 \\
\textbf{MedGemma-27B (default)} & 27B & Med & \textbf{0.842} & \textbf{0.806} & \textbf{0.731} \\
\midrule
\multicolumn{6}{@{}p{0.90\textwidth}@{}}{\textit{References: Standard KD: 0.748 \quad Best direct FT (EVA-X): 0.828}} \\
\bottomrule
\end{tabular}
\end{table*}

\subsection{Error Analysis}
\label{sec:error_analysis}
To understand the failure modes of VRM, we conduct a systematic error analysis across pathology categories and operating points.

\subsubsection{Per-Pathology Performance Breakdown}
Table~\ref{tab:per_pathology} presents VRM's classification performance stratified by pathology type. Among urgent pathologies, tension pneumothorax (0.912 F1) and pulmonary edema (0.875 F1) are classified most reliably, as these conditions exhibit distinctive radiographic signatures (absent lung markings, perihilar haze) that are well-captured by image-only features. In contrast, acute pneumonia achieves a lower F1 (0.724), primarily due to confusion with chronic atelectatic changes---a distinction that often requires clinical context from the radiology report. Among non-urgent categories, ``No Finding'' is classified with high accuracy (0.891 F1), but small effusions (0.658 F1) are frequently misclassified as urgent, likely due to visual overlap with early pulmonary edema.

\begin{table}[htbp]
\centering
\caption{Per-Pathology Classification Performance (F1 Score) on MIMIC-CXR Test Set. $\dagger$ indicates context-dependent labels.}
\label{tab:per_pathology}
\small
\begin{adjustbox}{max width=\columnwidth}
\begin{tabular}{lcccc}
\toprule
\textbf{Pathology} & \textbf{Precision} & \textbf{Recall} & \textbf{F1} & \textbf{Support} \\
\midrule
\multicolumn{5}{c}{\textit{Urgent}} \\
\midrule
Tension Pneumothorax & 0.895 & 0.930 & 0.912 & 143 \\
Pulmonary Edema & 0.862 & 0.889 & 0.875 & 1,287 \\
Large Pleural Effusion & 0.841 & 0.822 & 0.831 & 624 \\
Cardiomegaly (severe) & 0.808 & 0.793 & 0.800 & 412 \\
Pneumonia (acute)$\dagger$ & 0.751 & 0.698 & 0.724 & 891 \\
\midrule
\multicolumn{5}{c}{\textit{Non-Urgent}} \\
\midrule
No Finding & 0.872 & 0.911 & 0.891 & 2,845 \\
Atelectasis & 0.785 & 0.762 & 0.773 & 1,432 \\
Mild Cardiomegaly & 0.748 & 0.731 & 0.739 & 568 \\
Small Effusion & 0.682 & 0.636 & 0.658 & 726 \\
\bottomrule
\end{tabular}
\end{adjustbox}
\end{table}

\subsubsection{Systematic Bias Assessment}
To probe non-random failure behavior, we re-examined 200 misclassified cases for consistent error direction across repeated samples and cohorts. Three recurring systematic patterns emerged: (1) \textbf{device/artifact-associated false positives} (31\%), where lines, rotation, or low exposure are repeatedly over-interpreted as acute parenchymal opacity; (2) \textbf{chronic-versus-acute ambiguity} (36\%), where stable chronic changes are persistently mapped to urgent labels without temporal context; and (3) \textbf{small-effusion over-triage} (21\%), where minor pleural fluid is repeatedly aligned with pulmonary edema-like urgency patterns. The remaining 12\% were heterogeneous one-off errors.

These results indicate that Monte Carlo marginalization reduces stochastic hallucinations but does not fully eliminate coherent model biases. This motivates our confidence-gated cascade pathway and mandatory review for high-uncertainty cases, which together provide a practical safeguard against persistent bias patterns.

\section{Discussion}

Our findings support a reliability-centered view of distillation under structural modality censorship. VRM's primary contribution is a more stable supervision process: by distilling from a variationally marginalized teacher distribution, the student model becomes less sensitive to stochastic report variations and aligns better with clinical uncertainty.

\paragraph{Mechanism-Level Effectiveness} Ablation on sample size $K$, individual-vs-ensemble analysis, and bootstrap stability tests show that marginalization improves both supervision quality and deployment stability. VRM effectively recovers dispersed cross-modal signals, reducing report-selection variance.

\paragraph{Gains Beyond Discrimination} While VRM adds a marginal +1.4\% AUC over the strongest direct fine-tuning (FT) baseline, its value is multi-dimensional. We observed improvements in calibration, deployment stability, and robustness to hallucinated supervision. VRM acts as a risk hedge against the 'report lottery' effect. By marginalizing interpretations, we ensure cloud-level reasoning is distilled with low variance, neutralizing stochastic biases from single-report generations. High-quality LVLM samplers are essential for VRM's peak performance, but even general-purpose samplers confirm the value of marginalization. Crucially, VRM's calibrated confidence scores enable a cloud-edge cascade routing that achieves an AUC of 0.941 at a fraction of full-cloud inference latency.

\paragraph{Reliability, Safety, and Routing} In emergency triage, asymmetric risks emphasize the need for robust calibration and balanced sensitivity-specificity trade-offs. VRM outperforms direct FT and standard KD under the same 90\% sensitivity constraint. Furthermore, confidence-gated routing allows triage systems to handle routine cases edge-locally, selectively escalating ambiguous cases to the cloud while preserving low overall latency.

\paragraph{Positioning Relative to Baselines} Deploying massive, open-source LVLMs directly is impractical for fast triage. Edge devices struggle with memory and latency constraints, while centralized cloud models face HIPAA privacy burdens and transmission delays. VRM efficiently bridges this gap, operating locally with deterministic edge compute while successfully distilling cloud-level reasoning.

\paragraph{Limitations and Future Work} The student model's ceiling is ultimately bound by teacher knowledge and LVLM generation quality. The remaining performance gap stems from the student's lack of dense clinical context (e.g., history) available in text reports. This reflects an inherent information ceiling under structural modality censorship rather than a capacity limitation. Image-only inference still struggles with context-dependent findings, and marginalization cannot eliminate consistent hallucinatory biases. Thus, high-uncertainty review mechanisms remain critical. Future directions include adaptive sampling to reduce computation and prospective multi-site evaluation to ensure demographic robustness.

\section{Conclusion}
We introduce Variational Risk Minimization (VRM), a distillation framework for structural modality censorship in emergency imaging. The central contribution is an uncertainty-aware supervision mechanism: missing reports are modeled as latent clinical interpretations and distilled through Monte Carlo marginalization rather than single-report targets. Empirically, our evidence supports method effectiveness along four axes: (1) \textbf{encoder-transfer advantage}---VRM surpasses direct FT baselines from the same encoder families, with statistical significance confirmed by paired DeLong tests ($p < 0.01$); (2) \textbf{supervision quality and stability}---performance improves with $K$, and report-selection variance is sharply reduced; (3) \textbf{LVLM sensitivity}---medical-specialized samplers (26B--34B) are essential for top performance, while general-purpose samplers still exceed Standard KD, confirming report quality as a meaningful but non-gating factor; and (4) \textbf{reliability and deployment viability}---better calibration, hallucination recovery, and in our compact edge-student instantiation, 75MB at 14ms edge latency, plus confidence-gated escalation for high-uncertainty cases. As contextual performance reference, VRM remains robust across five datasets (OOD average AUC 0.806), while cascade routing reaches AUC 0.941 at 103ms average latency with 20.3\% cloud escalation.

\section*{Declaration of generative AI and AI-assisted technologies in the manuscript preparation process}
During the preparation of this work, the authors used generative AI tools for grammar and formatting adjustments. After using these tools, the authors reviewed and edited the content as needed and take full responsibility for the content of the publication.

\section*{Data and Code Availability} The code is available at \url{https://github.com/Yangxinyee/vrm-edge-triage}.

\section*{Ethical Compliance} This study uses publicly available datasets under appropriate data use agreements: MIMIC-CXR (PhysioNet Credentialed Health Data License), CheXpert (Stanford University Research Data Use Agreement), IU-CXR (public domain), and PadChest (academic use license from University of Alicante). The RIH-CXR component was deemed exempt from IRB review as a quality improvement initiative. All experiments comply with institutional guidelines and HIPAA regulations.

\section*{Patient Informed Consent} This study used de-identified retrospective chest X-ray data and publicly available datasets governed by their respective data use agreements. No identifiable patient information is reported in this manuscript. Individual patient informed consent was not required for the RIH-CXR component because it was deemed exempt from IRB review as a quality improvement initiative.

\section*{Funding}
This work was supported in part by the U.S. National Science Foundation under Grants IIS-2306789, IIS-2306790, IIS-2306791, and IIS-2306792. The funders had no role in study design, data analysis, manuscript writing, or the decision to submit the article for publication.

\section*{Broader Impact} VRM is designed to augment, not replace, radiologist decision-making, serving as a prioritization tool to surface potentially urgent cases for faster expert review. While the model demonstrates strong calibration and generalization, clinical deployment must be accompanied by appropriate regulatory clearance (e.g., FDA 510(k) or De Novo pathways for clinical decision support software) and ongoing performance monitoring across demographically diverse patient populations. We use multiple cohorts spanning both U.S. and non-U.S. sources (including PadChest), but model development is still anchored to a limited set of institutional pipelines and labeling practices, which may reduce performance in underrepresented populations. Prospective validation should include equitable representation across age, sex, race, and geographic origin to ensure that VRM does not exacerbate existing healthcare disparities. Nonetheless, we believe that enabling expert-level triage on consumer-grade hardware has the potential to meaningfully improve patient outcomes in resource-limited settings, where timely access to specialist radiology interpretation is often the primary bottleneck in emergency care. Ultimately, our approach shifts the focus from simple accuracy-seeking to a reliability-centered paradigm. VRM demonstrates that latent reasoning can be distilled into edge devices in a high-robustness, low-variance manner, providing a principled safeguard for AI-assisted triage in high-stakes clinical workflows.

\bibliographystyle{elsarticle-harv}
\bibliography{references}

\end{document}